\documentclass[sigconf]{acmart}
\usepackage{booktabs} % For formal tables

\usepackage[ruled]{algorithm2e} % For algorithms
\usepackage{amssymb}
\usepackage{color}
\usepackage{todonotes}
\usepackage{listings}
\usepackage{amsmath}
\usepackage{booktabs}
\usepackage{enumitem}
\usepackage{algpseudocode}
\usepackage{multirow}
\usepackage{url}
\usepackage{graphicx}
\usepackage{subfig}
\usepackage{balance}
\usepackage{adjustbox}
\usepackage{tabularx,ragged2e,booktabs}

\newcommand{\uj}[1]{\textcolor{cyan}{\textbf{}}}
\newcommand{\sd}[1]{\textcolor{green}{\textbf{}}}
\newcommand{\mr}[1]{\textcolor{red}{\textbf{}}}

\begin{document}
% Title portion
\title{Predicting User Knowledge Gain \\ in Informational Search Sessions} 

\author{Ran Yu}
\affiliation{%
  \institution{L3S Research Center}
  \streetaddress{Appelstr. 4}
  \city{Hannover} 
  \country{Germany} 
  \postcode{30167}
}
\email{yu@l3s.de}

\author{Ujwal Gadiraju}
\affiliation{%
  \institution{L3S Research Center}
  \streetaddress{P.O. Box 1212}
  \city{Hannover} 
  \state{Germany} 
  \postcode{30167}
}
\email{gadiraju@l3s.de}

\author{Peter Holtz}
\affiliation{%
  \institution{Leibinz Insitut f\"ur Wissensmedien}
  \streetaddress{P.O. Box 1212}
  \city{T\"ubingen} 
  \state{Germany} 
  \postcode{72076}
}
\email{holtz@iwm-tuebingen.de}

\author{Markus Rokicki}
\affiliation{%
  \institution{L3S Research Center}
  \streetaddress{P.O. Box 1212}
  \city{Hannover} 
  \state{Germany} 
  \postcode{30167}
}
\email{rokicki@l3s.de}

\author{Philipp Kemkes}
\affiliation{%
  \institution{L3S Research Center}
  \streetaddress{P.O. Box 1212}
  \city{Hannover} 
  \state{Germany} 
  \postcode{30167}
}
\email{kemkes@l3s.de}

\author{Stefan Dietze}
\affiliation{
  \institution{L3S Research Center}
  \streetaddress{Appelstr. 4}
  \city{Hannover} 
  \country{Germany} 
  \postcode{30167}
}
\email{dietze@l3s.de}

\begin{abstract}

Web search is frequently used by people to acquire new knowledge and to satisfy learning-related objectives. In this context, informational search missions with an intention to obtain knowledge pertaining to a topic are prominent. The importance of learning as an outcome of web search has been recognized. Yet, there is a lack of understanding of the impact of web search on a user's knowledge state. Predicting the knowledge gain of users can be an important step forward if web search engines that are currently optimized for relevance can be molded to serve learning outcomes. 
In this paper, we introduce a supervised model to predict a user's knowledge state and knowledge gain from features captured during the search sessions. To measure and predict the knowledge gain of users in informational search sessions, we recruited 468 distinct users using crowdsourcing and orchestrated real-world search sessions spanning 11 different topics and information needs. By using scientifically formulated knowledge tests, we calibrated the knowledge of users before and after their search sessions, quantifying their knowledge gain. Our supervised models utilise and derive a comprehensive set of features from the current state of the art and compare performance of a range of feature sets and feature selection strategies. Through our results, we demonstrate the ability to predict and classify the knowledge state and gain using features obtained during search sessions, exhibiting superior performance to an existing baseline in the knowledge state prediction task.

\end{abstract}



%\keywords{search as learning; knowledge gain; web search; user modeling}

\thanks{}

\copyrightyear{2018} 
\acmYear{2018} 
\setcopyright{acmlicensed}
\acmConference[SIGIR '18]{The 41st International ACM SIGIR Conference on Research and Development in Information Retrieval}{July 8--12, 2018}{Ann Arbor, MI, USA}
\acmBooktitle{SIGIR '18: The 41st International ACM SIGIR Conference on Research and Development in Information Retrieval, July 8--12, 2018, Ann Arbor, MI, USA}
\acmPrice{15.00}
\acmDOI{10.1145/3209978.3210064}
\acmISBN{978-1-4503-5657-2/18/07}

\maketitle

%\renewcommand{\shortauthors}{G. Zhou et al.}
\vspace{-10pt}
\section{Introduction}

Searching the web for information is among the most frequent online activities. Broder categorized web search queries into having either \textit{navigational}, \textit{transactional} or \textit{informational} intents \cite{broder2002taxonomy}. In informational web search sessions, the intent of a user is to acquire some information assumed to be present on one or more web pages.

Recent research in the \textit{search as learning} (SAL) domain has recognized the importance of learning scopes and focused on observing and detecting learning needs during web search.
Eickhoff et al. investigated the correlation between several query and search mission-related metrics and learning progress \cite{eickhoff2014lessons}. Wu et al. predicted the  difficulty of search tasks from query and mission-related features \cite{wu2012grannies}. Collins-Thompson et al. investigated the effectiveness of user interaction with respect to certain learning outcomes \cite{collins2016assessing}. In addition, \cite{zhang2011predicting} has shown that data obtained during the search process provides valuable indicators about the domain knowledge of a user. 

Although the importance of learning as an implicit element of web search has been established, there is still only a limited understanding of the impact of search behavior on a user's knowledge state and knowledge gain.  Prior work has focused on improving the learning experience and efficiency during search sessions, but the measurement of a user's knowledge gain through the course of an informational search session has not yet been addressed. This is in part due to the difficulty in accurately quantifying knowledge gain through the course of a search session. If web search engines that are currently optimized for relevance can be re-molded to serve learning outcomes, the capability to predict knowledge gain will be a crucial step forward.

In this paper, we aim to address the aforementioned gap. We used crowdsourcing to recruit users who participated in real-world search sessions spanning 11 different topics and information needs. By using scientifically formulated knowledge tests, we calibrated the knowledge of users before and after their search sessions, quantifying their knowledge gain. We introduce a supervised model to predict a user's knowledge state and knowledge gain from features captured during the search sessions. 

\noindent\textbf{Original Contributions.}
Through our work in this paper, we make the following contributions to the current body of literature:
\begin{itemize}[leftmargin=*, nosep]
\item A model for predicting the user's knowledge gain and state during real-world informational search sessions. 
\item An analysis of the affect of user interactions (ranging from the queries entered to their browsing behavior) on their knowledge state and knowledge gain.
\item We release a dataset capturing user behavior and interactions in 468 experimentally orchestrated informational search missions, capturing all features across the aforementioned dimensions as well as knowledge assessments obtained through pre- and post-tests. Given the lack of comparable datasets which are both recent and publicly available, we anticipate that this corpus can facilitate SAL research related to different tasks.
\end{itemize}

\noindent\textbf{Implications.} 
The capability to predict a user's knowledge state and gain through the course of an informational search session has the potential to reshape search engines to support learning outcomes as an implicit part of retrieval and ranking. This is of particular importance given that Web search already augments learning processes in a variety of informal as well as formal learning scenarios, such as classrooms, libraries and in work environments. Our contributions advance the current understanding of learning through web search, setting important precedents for further research.

\section{Related Work} \label{sec:sota}
We discuss two main realms of closely related work -- studies on the relation between (i) a user's search behavior and knowledge gain, and (ii) a user's search behavior and knowledge state.

\subsection{Search Behavior and Knowledge Gain}
Eickhoff et al.~\cite{eickhoff2014lessons} investigated the correlation between a number of features extracted from search session as well as SERP (Search Engine Results Page) documents with learning needs related to either procedural or declarative knowledge. 
Results obtained from an analysis of large-scale query logs showed the distinct evolution of particular features throughout search sessions and the correlation of document features with the actual learning intent. 
The influence of distinct query types on knowledge gain was studied by Collins-Thompson et al.~\cite{collins2016assessing}, finding that intrinsically diverse queries lead to increased knowledge gain. 
Gadiraju et al.~\cite{gadiraju2018chiir} described the use of knowledge tests to calibrate the knowledge of users before and after their search sessions, quantifying their knowledge gain. They investigated the impact of information needs on the search behavior and knowledge gain of users.

Studies on exploratory search have also investigated a similar set of search behaviors that influence the learning outcome. Hagen et al.~\cite{hagen2016writers} investigated the relation between the writing behavior and the exploratory search pattern of writers. 
The authors revealed that query terms can be learned while searching and reading. 
In addition, Vakkari~\cite{vakkari2016searching} provided a structured survey of features indicating learning needs as well as user knowledge and knowledge gain throughout the search process. Zhuang et al.~\cite{zhuang2017understanding} investigated the possibility of using 37 user search behavioral features to predict the user engagement with supervised classifiers. As the engagement in the search process usually is correlated with learning outcome, in our work we have also taken into consideration the set of features have been studied in this work.

The aforementioned prior works have either studied a limited set of features or have addressed only specific learning scenarios and learning types. The generalizability of knowledge gain measures in previous works has not been investigated. In this paper, we extend the current understanding of user knowledge gain in informational search sessions. Using real world information needs and search sessions on the Web, we investigate the possibility of using search activity related features to predict knowledge gain.

\subsection{Search Behavior and Knowledge State}

By matching the learning tasks into different learning stages of Anderson and Krathwohl's taxonomy~\cite{anderson2001taxonomy}, Jansen et al. studied the correlation between search behaviors of 72 participants and their learning stage \cite{jansen2009using}. They showed that information searching is a learning process with unique searching characteristics corresponding to particular learning levels. 
Gwizdka et al.~\cite{gwizdka2016towards} proposed to assess learning outcomes in search environments by correlating individual search behaviors with corresponding eye-tracking measures. 
Syed and Collins-Thompson~\cite{syed2017retrieval} proposed to optimize the learning outcome of the vocabulary learning task by selecting a set of documents while considering keyword density and domain knowledge of the learner.

White et al.~\cite{white2009characterizing} investigated the difference between the behavior of domain experts and non-experts in seeking information on the same topic. By analyzing the activity log of experts and non-experts across different domains, the authors found that the distribution of features such as number of queries and query length differed across the levels of expertise. Zhang et al. ~\cite{zhang2015predicting, zhang2011predicting} explored using search behavior as an indicator for the domain knowledge of a user. Through a small study ($n=35$), they identified features such as the average query length or the rank of documents consumed from the search results as being predictive. Further, Cole et al.~\cite{cole2013inferring}, observed that behavioral patterns provide reliable indicators about the domain knowledge of a user, even if the actual content or topics of queries and documents are disregarded entirely.

Other studies have focused on detecting task difficulty in search environments based on user activity data in situations where the subjective assessment of task difficulty is highly correlated to the user's domain knowledge~\cite{li2008faceted,gwizdka2006can}. 
Gwizdka and Spence~\cite{gwizdka2006can} showed that a searcher's perception of task difficulty is a subjective factor that depends on the domain knowledge and some other individual traits. 
Arguello~\cite{arguello2014predicting} proposed to use logistic regression to predict task difficulty in a search environment. Data was collected through a crowdsourcing platform, and the author used search tasks created by Wu et al.~\cite{wu2012grannies}, which contain task difficulty assessments on multiple dimensions. 

The aforementioned studies focused on investigating the relation between search behavior and a user's knowledge state. 
Our work leverages these results to derive a comprehensive feature set for our supervised models.
In contrast to prior works, we aim at \emph{predicting} the knowledge state of a user -- avoiding the need for explicit post-search knowledge assessments.

\section{Problem Definition} \label{sec:problem}

In the context of Web search, Broder~\cite{broder2002taxonomy} classified search queries according to their intent into three classes: 1) navigational, 2) informational, and 3) transactional. Herein, \textit{informational queries} are defined as those queries where `the intent of a user is to acquire some information assumed to be present on one or more web pages'~\cite{broder2002taxonomy}. 
Thus, \textit{informational queries} imply a particular learning intent; \textit{intentional learning} is generally defined as learning that is motivated by intentions and is goal directed~\cite{blumschein2012intentional}, in contrast to latent or incidental learning. 

Based on the constructs of intentional learning and informational queries, we arrive at the following definition:

\begin{definition}{\textit{Intentional Learning-Related Search Session.}}
An intentional learning-related \textit{search session} comprises of the sequence of a user's actions, with respect to satisfying her learning intent in a web search environment through informational queries. A user's sequence of actions begins with querying the web, and includes browsing through the search results, click and scroll activity, navigation via hyperlinks, query reformulations, and so forth.
\end{definition}

For the sake of simplicity, we henceforth refer to informational sessions, i.e. sessions with a particular learning intent, as ``sessions''.

In this paper, from the observed user interactions in informational search sessions, we aim to predict (i) the \textit{knowledge state} and (ii) \textit{knowledge gain} of a user as follows. 

\begin{definition}{\textit{Predicting a User's Knowledge State and Gain During Search Sessions.}}
Let $s$ be a search session starting at time $t_i$ and ending at time $t_j$ aimed at satisfying a particular information need, that is, a learning intent $\iota$ of user $u$. 
Based on the user interactions during session $s$ captured in the time period $[t_i, t_j]$, we aim to:
\begin{enumerate}[leftmargin=*, nosep]
\item classify the knowledge state (KS) $k(t_j)$ of $u$ at time point $t_j$ with respect to a particular information need. For the sake of this work, a user's knowledge state with respect to a particular information need is defined by the user's capability to correctly respond to a set of  questions about the corresponding information need. We classify a user's knowledge state into 3 classes according to her capability: low knowledge state, moderate knowledge state and high knowledge state (Section \ref{sec:data_grouping}). 

\item classify the knowledge state change, i.e. the knowledge gain (KG) $\Delta k(t_i, t_j)$ of $u$ during time period $[t_i, t_j]$ into different degrees. Similarly, a user's knowledge gain with respect to a particular information need is defined as the improvement of user capability (accuracy) to correctly respond to a set of test questions about the corresponding information need. We classify user knowledge gain into 3 classes according to the improvement of user capability: low knowledge gain, moderate knowledge gain and high knowledge gain (Section \ref{sec:data_grouping}). 
\end{enumerate}

\end{definition}

\section{Obtaining Search Session Data} \label{sec:data}
We adopted a crowdsourcing approach and orchestrated search sessions with varying information needs. All interactions of the users during the search sessions were logged. We analyzed the data to further the understanding of user knowledge evolution in informational search sessions on the Web. In this section, we describe the study design and experimental setup.

\subsection{Study Design and Search Environment}
\label{subsec:study}

We recruited participants from CrowdFlower\footnote{\url{http://www.crowdflower.com/}}, a premier crowdsourcing platform. At the onset, workers were informed that the task entailed `searching the Web for some information'. Workers willing to participate were redirected to our external platform, \textit{SearchWell}\footnote{http://searchwell.l3s.uni-hannover.de/?uid=12345678.}, a search system built on top of the Bing Web Search API. We logged worker activity on the platform including mouse movements, clicks, and key presses, using PHP/Javascript and the jQuery library. 
Workers were first asked to respond to a few questions (called `\textit{items}') corresponding to a particular topic without searching the Web for answers. The questions took the form of statements pertaining to a topic, and workers had to select whether the statement was `TRUE', `FALSE', or `I DON'T KNOW' in case they were not sure. In this way, we calibrated the knowledge of users corresponding to a given topic. To encourage the workers to respond without external consultation, we informed them that their responses to these questions would not affect their pay. We also encouraged workers to avoid guessing. The results of this pre-test were used to calibrate the knowledge of the workers with respect to the topic. We describe the topics and how the knowledge tests were created in the following Section \ref{subsec:info_need}. On completing the knowledge calibration test, workers were presented with their actual task. 

Workers were presented an \textit{information need} corresponding to the topic of the calibration test they completed. They were told to use the \textit{SearchWell} platform to search the Web and satisfy their information need. To incentivize workers towards realistic attempts to learn about the topic, we informed them that they will have to complete a final test on the topic to successfully finish the task. Furthermore, workers were conveyed the message that depending on their accuracy on the final test they could earn a bonus payment. We subsequently logged all the activities of the workers (mouse movements, key presses and clicks) within the \textit{SearchWell} platform. Workers were allowed to begin the final test anytime after a  
search session, which is when a link to the final test was made available. Workers were encouraged to proceed to the next stage only once they felt that their information need was satisfied and when they were ready for the post-session test. On completing the post-session test, workers received a unique code that they could enter on CrowdFlower to claim their reward.

We restricted the participation to workers from English-speaking countries to ensure that they understood the task and instructions adequately \cite{gadiraju2015understanding,gadirajuclarity}. To ensure reliability of the resulting data, we restricted the participation to \textit{Level-3 workers\footnote{\textit{Level-3 contributors} on CrowdFlower comprise workers who completed over $100$ test questions across hundreds of different types of tasks, and have a near perfect overall accuracy. They are workers of the highest quality on CrowdFlower.}} on CrowdFlower.

\subsection{Topics -- Defining Information Needs}
\label{subsec:info_need}

\begin{table*}[ht]
	\caption{Topics and corresponding information needs presented to participants in the informational search sessions, along with the internal reliability of the corresponding knowledge tests. `$\alpha$1', `$\alpha$2' represent Cronbach's $\alpha$ for the pre-session test and post-session test respectively. `N' is the number of reliable participants after filtering.}
    \vspace{-10pt}
	\label{tab:topics}
	\small
	\centering
	\scalebox{.9}{
	\begin{tabular}{p{3.6cm}p{14cm}p{0.25cm}p{0.25cm}p{0.2cm}} 
	\toprule
	\textbf{Topic} & \textbf{Information Need}& \textbf{$\alpha$1}& \textbf{$\alpha$2} & 
    \textbf{N}\\
	\midrule
    
	{1. Altitude Sickness} & {In this task you are required to acquire knowledge about the symptoms, causes and prevention of altitude sickness. (20 items)}&0.59 &0.79 & 44 \\

	{2. American Revolutionary War} & {In this task, you are required to acquire knowledge about the `American Revolutionary War'. (10 items)} &0.74 &0.55 & 39\\

	{3. Carpenter Bees} & {In this task, you are required to acquire knowledge about the biological species `carpenter bees'. How do they look? How do they live? (10 items)} &0.79 &0.58 & 43 \\

	{4. Evolution} & {In this task, you are required to acquire knowledge about the theory of evolution. (12 items)}&0.55 &0.72 & 42 \\

	{5. NASA Interplanetary Missions} & {In this task, you are required to acquire knowledge about the past, present, and possible future of interplanetary missions that are planned by the NASA. (20 items)} &0.80 &0.75 & 40\\ 
    
	{6. Orcas Island} & {In this task you are required to acquire knowledge about the Orcas Island. (20 items)} &0.91 &0.85 & 38\\
    
	{7. Sangre de Cristo Mountains} & {In this task, you are required to acquire knowledge about `Sangre de Cristo' mountain range. (10 items)} &0.70 &0.52 & 38\\
    
	{8. Sun Tzu} & {In this task, you are required to acquire knowledge about the Chinese author Sun Tzu - about his life, his writings, and his influence to the present day. (15 items)} &0.81 &0.63 & 35\\
    
	{9. Tornado} & {In this task, you are required to acquire knowledge about the weather phenomenon that is called `tornado' (20 items)}&0.82 &0.62  & 37 \\

	{10. USS Cole Bombing} & {In this task, you are required to acquire knowledge about the 2000 terrorist attack that came to be known as the `USS Cole bombing'. (10 items)}&0.83 &0.55 & 38 \\

	{11. HIV} & {In this task you are required to acquire knowledge about the transmission, prevention, and consequences of HIV infection. (45 items)} &0.87 &0.84 & 74\\
	\bottomrule
	\end{tabular}}
    \vspace{-10pt}
\end{table*}

We constructed a corpus of topics representing varying scopes of information needs (with some relatively broader than others). Topics were selected 
from the \textit{TREC 2014 Web Track} dataset\footnote{\url{http://www.trec.nist.gov/act_part/tracks/web/web2014.topics.txt}}, and corresponding information needs were defined accordingly. In all cases, the knowledge of users before beginning an informational search session was assessed using pre-tested and evaluated \textit{knowledge tests}. Knowledge tests are scientifically formulated tests that measure the knowledge of a participant on a given topic (for example, the HIV knowledge test \cite{carey1997hiv}). Topics and test items are available online\footnote{https://sites.google.com/view/predicting-user-knowledge \label{fn:dataset}}.

Knowledge on all given topics was measured using knowledge tests comprising of between 10 and 45 items. The answer options were in all cases `TRUE', `FALSE', and `I DON'T KNOW'. The differences in the number of items reflects our attempt to feature varying scopes of information needs; relatively narrow (e.g., \textit{Carpenter Bees}--10 items) as well as broad (e.g., \textit{NASA Interplanetary Missions}--20 items). In the construction of all scales, an item pool comprising of more items than finally used was constructed. After a pilot test with 100 distinct participants recruited via CrowdFlower for each of the 11 topics, items that proved to be either too easy (e.g., more than 80\% correct answers) or too hard/ambiguous (e.g., more false than true answers) were discarded. Table \ref{tab:topics} presents the topics and corresponding information needs considered for orchestrating the informational search sessions. It also shows the internal reliability (using Cronbach's $\alpha$) of the pre- and post-session knowledge tests corresponding to each topic. We observe moderate to high values of $\alpha$ in the pre- and post session knowledge tests, suggesting a desirable level of internal consistency. 

\subsection{Data Collection}
\label{subsec:searchwell}

To further ensure the reliability of responses and the behavioral data thus produced in the search sessions, we filtered workers using the following criteria.

\begin{itemize}[leftmargin=*, nosep]
\item Workers who entered no queries in the \textit{SearchWell} system. Since the aim of our work is to further the understanding of how the knowledge state of a user evolves in informational search sessions, we discard those users who did not enter a search query. 
\item Workers who selected the same option; either `YES' or `NO', for all items in the calibration test or the post-session test.
\item Workers who did not complete the post-session test.
\end{itemize}

We filtered out 132 workers due to the aforementioned criteria, resulting in 468 workers across the 11 topics. The analysis and results presented hereafter are based on these 468 sessions alone. For the benefit of further research in this community, the filtered data has been thoroughly anonymized and made publicly available\textsuperscript{\ref{fn:dataset}}. 
We henceforth refer to these filtered workers as users in our experimentally orchestrated information search sessions.

\subsection{Descriptive Analysis -- Important Details}

We measure the knowledge gain of users in search sessions corresponding to an information need as the difference between their knowledge calibration score and the post-session test score\footnote{We consider the `I DON'T KNOW' options that were selected, as incorrect responses while computing the knowledge calibration scores and post-session test scores.}. Table \ref{tab:topicVSkg} presents the average knowledge calibration scores, post-session test scores, and the resulting knowledge gain of users across search sessions with different information needs. Across all topics and search sessions, we found that users exhibited an average knowledge gain of around 19\%. Nearly 70\% of all the workers exhibited a knowledge gain, while the remaining workers did not. The standard deviation observed in the knowledge gain of users across all topics is notably high, due to the varying domain knowledge of users. This is evident from the average calibration scores in Table \ref{tab:topicVSkg}.

\begin{table}[h]
\centering
\caption{The average knowledge gain of users across the different topics. To enhance readability, the rows have been ordered by ascending knowledge gain.}
\vspace{-10pt}
\scalebox{.75}{
	\begin{tabular}{lccc}
	\toprule
	\textbf{Topic /}	&	\textbf{Avg. Calibration}	&	\textbf{Avg. Post}	&	\textbf{Knowledge Gain}	\\
   \textbf{Information Need} 		&	\textbf{Score} \textit{(in \%)}	&	\textbf{Score} \textit{(in \%)}	&	\textit{(in \%)}	\\	
    \midrule

HIV \textit{(N=74)}	&	$66.25\pm14.86$	&	$71.68\pm14.48$	&	$5.44\pm10.02$	\\
Evolution \textit{(N=42)}	&	$34.07\pm17.99$	&	$48.15\pm22.49$	&	$14.07\pm18.66$	\\
NASA Interplanetary 	&	$38.1\pm20.53$	&	$52.5\pm17.43$	&	$14.40\pm22.10$	\\
Missions \textit{(N=40)}	&		&		&		\\
Altitude Sickness \textit{(N=44)}	&	$55.88\pm16.31$	&	$70.66\pm19.11$	&	$14.78\pm17.76$	\\
Sangre de Cristo	&	$33.25\pm22.40$	&	$49.75\pm18.10$	&	$16.50\pm22.31$	\\
Mountains \textit{(N=38)}	&		&		&		\\
Tornados \textit{(N=37)} 	&	$34.44\pm21.02$	&	$53.47\pm16.28$	&	$19.03\pm22.010$	\\
Sun Tzu \textit{(N=35)}	&	$40.54\pm23.37$	&	$60.18\pm17.15$	&	$19.64\pm21.59$	\\
American Revolutionary &	$34.52\pm25.65$	&	$55.95\pm20.71$	&	$21.43\pm27.31$	\\
War \textit{(N=39)}	&		&		&		\\
Carpenter Bees \textit{(N=43)} 	&	$45.65\pm27.08$	&	$67.17\pm20.29$	&	$21.52\pm30.50$	\\
USS Cole Bombing \textit{(N=38)}	&	$30.95\pm25.22$	&	$54.37\pm16.29$	&	$23.41\pm31.30$	\\
Orcas Island \textit{(N=38)}	&	$34.74\pm30.08$	&	$65.51\pm22.04$	&	$30.77\pm30.25$	\\
    \midrule
\textbf{Overall} \textit{(N=468)}	&	$40.76\pm22.23$	&	$59.04\pm18.58$	&	$18.27\pm23.07$	\\
	\bottomrule
	\end{tabular}}
	\label{tab:topicVSkg}
    \vspace{-5pt}
\end{table}

We found that on average, the highest knowledge gain was observed through the search sessions corresponding to the topic, `\textit{Orcas Island}', while the least knowledge gain was observed through those corresponding to the topic, `\textit{HIV}'. We found a negative linear relationship between topics that workers were familiar with (indicated by their calibration scores) and their knowledge gain; \textbf{R}$=-.65, R^2=.42, p<.001$. This suggests that the more popular a topic is, or the more familiar that users are with a topic, the lesser they tend to learn about the topic in informational search sessions. Thus, we found that 42\% of the variance in the knowledge gain of users can be explained by the topic familiarity.

We found that the average session length of users across the different topics was nearly 5 mins long (\textit{M=4.82, SD=5.20}). During the search sessions, users navigated to over 5 web pages on average (\textit{M=5.46, SD=3.41}). We note that on average users entered 2 distinct queries in a search session (\textit{M=2.20, SD=2.18}), with an average query length of just over 4 terms (\textit{M=4.56, SD=2.63}). For each query that was entered, users navigated to over 3 web pages on average (\textit{M=3.27, SD=1.60}). Users spent nearly 2 minutes actively on web pages they navigated to (\textit{M=1.97, SD=1.52}). 

\subsection{Knowledge State and Knowledge Gain Classes}\label{sec:data_grouping}
\begin{table}[h]
\vspace{-5pt}
\centering
\caption{\small{User groups created based on $average \pm 0.5SD$.}}
\vspace{-10pt}
%\scalebox{0.75}{
\begin{adjustbox}{max width=0.35\textwidth}
\small
\begin{tabular}{ p{0.5cm} p{0.7cm}  p{0.7cm}p{1cm} p{1.2cm}p{1cm}}
\toprule
\textbf{Task}& \textbf{Mean }& \textbf{SD} & \textbf{Low} &\textbf{Moderate}& \textbf{High}\\ 
\midrule
\textbf{KG} & 0.193 & 0.231 & 167&179&122\\
\textbf{KS}& 0.618 & 0.191 &145&171&152 \\
\bottomrule 
\end{tabular}
\end{adjustbox}
\label{tab:user_group}
\vspace{-10pt}
\end{table}

We used a \textit{Standard Deviation Classification} approach to obtain three classes of learners with regard to their level of knowledge. Assuming approximately normal distributions of the respective test scores (X) for the different topics, we transformed the test scores into Z-scores with a mean of 0 and a Standard Deviation (SD) of 1 (standardization). We then used statistically defined intervals (X < 0.5 SD = low; -0.5 SD < X < 0.5 SD = moderate; 0.5 SD < X = high) for the classification of the learners into roughly equal groups with low, moderate, or high knowledge. The same procedure was repeated for knowledge gain. Here as well, the empirical knowledge gain for every test was transformed into corresponding Z-scores and three roughly equal groups (low knowledge gain; moderate knowledge gain; high knowledge gain) were defined accordingly. 
In view of the substantial variety of different topics, we argue that such a tripartite categorization of knowledge states and knowledge gains respectively allows for the construction of robust models, which are themselves based on a large variety of features. Thus, insights from the learning tasks considered can be generalized  to other similar intentional learning activities. This procedure weighs all different knowledge tests equally irrespective of the number of items.

\section{Feature Extraction and Analysis}

We approach the problem of predicting knowledge state ($k(t_j)$) and knowledge gain ($\Delta k(t_i, t_j)$) described in Section \ref{sec:problem} with supervised models for classification, where details about the applied classification models are given in Section~\ref{sec:config}.  
To this end, each session $s$ is represented by a feature vector $ \vec{v} = (f_1, f_2, ..., f_n)$, where considered features are described in Section~\ref{sec:feature} and analyzed in Section~\ref{sec:feature_selection}. 

\subsection{Features Considered}
\label{sec:feature}

We extracted features according to multiple dimensions of a search session, structured into five categories, namely features related to the \textit{session}, \textit{queries}, \textit{SERP}, \textit{browsing} behavior and \textit{mouse} movements. 
The SERP category consists of features extracted from direct interactions with SERP items, while the browsing category consists of features extracted from subsequent user navigation beyond simple SERP clicks. The majority of features is motivated by existing literature, yet none of the features have been used on the inferential tasks of this work. 

All considered features $f_i$ are listed in Table~\ref{tab:features} together with the Pearson Correlation Coefficient scores $Corr(f_i, \Delta k(t_i, t_j) )$, $Corr(f_i, k(t_j) )$  between the respective feature and the knowledge gain (state).

\begin{table*}[h]
\centering
\caption{\small{Features for prediction of knowledge gain and knowledge state.}}
\vspace{-10pt}
\begin{adjustbox}{max width=0.98\textwidth}
\small
\begin{tabular}{ p{0.8cm} p{5.3cm} p{3.1cm}p{3.1cm} p{9cm}}
\toprule
\textbf{Category} & \textbf{Notation} & \textbf{$Corr(f_i, \Delta k(t_i, t_j) )$} & \textbf{$Corr(f_i, k(t_j) )$}& \textbf{Feature description} \\

\midrule

\multirow{2}{*}{\emph{Session}} & $s\_duration$ &-0.020&0.066& Duration of the search session of a worker on a given topic \\
& $s\_duration\_per\_q$ &-0.019&0.066& Session duration per query \\
\midrule

\multirow{6}{*}{\emph{Query}} & $q\_num$ &0.052&0.103& Number of queries in session $s$ \\[0.02cm]
& $q\_term\_\{max,min,avg,total\}$ &\{0.0002,-0.094,-0.042,0.047\}&\{0.065,0.032,0.051,0.068\}& Maximum, minimum, average, total number of query terms \\[0.02cm]
& $q\_uniq\_term\_\{max, min, avg, total\}$ &\{0.016,-0.087,-0.024,0.06\}&\{0.104,0.05,0.084,0.089\}&  Maximum, minimum, average number of unique terms per query \\[0.02cm]
& $q\_uniq\_term\_ratio$ &0.083&-0.002& Number of query terms / unique query terms ($\frac {q\_uniq\_term\_total}{q\_term\_total}$) \\[0.02cm]
& $q\_len\_\{first, last\}$ &\{-0.049,0.055\}&\{0.031,0.105\}& First, last query length \\[0.02cm]
& $q\_uniq\_term\_\{first,last\}$ &\{-0.023,-0.040\}&\{0.036,0.087\}& Number of unique terms of first, last query\\[0.02cm]
& $q\_complexity\_\{max,min,avg\}$ &\{0.097,0.086,0.093\}&\{0.087,0.078,0.049\}& Maximum, minimum, average of query complexity\\[0.02cm]
& $q\_complexity\_max\_diff$ &\{0.092\}&\{0.077\}& Difference between the maximum and minimum complexity\\[0.02cm]

\midrule

\multirow{6}{*}{\emph{SERP}}& $SERP\_click$ &-0.009&0.063&Total number of click on search result\\[0.02cm]
&$SERP\_click\_rank\_\{highest,lowest,avg\}$ &\{-0.101,-0.021,-0.017\}&\{-0.063,0.047,0.095\}& Average, highest, lowest rank of the clicks \\[0.02cm]
 & $SERP\_click\_interval$ &0.036&0.022& Average interval between clicks\\[0.02cm]
& $SERP\_click\_per\_query$ &-0.007&-0.012& Average number of clicks per query\\[0.02cm]
&$SERP\_no\_click\_query\_\{num,pct\}$&\{0.041,-0.051\}&\{0.077,0.029\}& Number, percentage of SERP with no clicks \\ [0.02cm]
& $SERP\_time\_\{total,avg,max\}$ &\{0.039,0.022,0.049\}&\{0.091,-0.008,0.043\}& Total, average, maximum time spend on SERPs \\[0.02cm]
& $SERP\_avg\_time\_to\_first\_click$ &-0.002&-0.027& Time till first click \\[0.02cm]
\midrule

\multirow{13}{*}{\emph{Browsing}} & $b\_num$ &-0.018&0.075& Total number of pages browsed in session\\[0.02cm]
& $b\_uniq\_num$ &0.029&0.109& Number of unique pages browsed in session \\[0.02cm]
& $b\_num\_per\_q $ &-0.017&-0.016& Average number of page browsed per query \\[0.02cm]
& $b\_uniq\_num\_per\_q $ &-0.017&-0.016& Average number of unique page viewed per query \\[0.02cm]
& $b\_time\_total$ &0.243&0.134& Total active time on the pages \\[0.02cm]
& $b\_time\_avg\_per\_q$ &0.236&0.063& Average active time on the browsed pages per query \\[0.02cm]
& $b\_time\_\{max,avg\}\_per\_page$ &\{0.306,0.291\}&\{0.104,0.089\}& Maximum, average active time on the browsed pages \\[0.02cm]
& $b\_revisited\_ratio$ &-0.058&-0.020& Ratio of revisited pages \\[0.02cm]
& $b\_\{num,pct\}\_from\_SERP $ &\{-0.017,0.058\}&\{0.074,0.056\}& Number, percentage of pages visited through SERP \\[0.02cm]
& $b\_\{num,pct\}\_from\_non\_SERP $ &\{-0.056,0.057\}&\{-0.028,0.025\}& Number, percentage of pages visited through pages other than SERP \\[0.02cm]
 &$b\_distinct\_domain\_num$  &-0.033&0.102& Number of distinct domains of the visited pages \\
& $b\_ttl\_len\_\{max,min,avg,total\}$  &\{-0.08,-0.058,-0.078,-0.109\}&\{0.146,0.106,0.146,0.082\}& Maximum, minimum, average, total page title length \\
& $b\_page\_size\_\{max,min,avg,total\}$  &\{0.109,-0.093,0.122,0.086\}&\{-0.055,-0.074,-0.057,-0.01\}& Maximum, minimum, average, total page size \\

&$b\_ttl\_q\_overlap\_\{max,min,avg,total\}$& \{0.15,0.089,0.14,0.091\} & \{0.005,-0.028,-0.018,0.023\}  & Maximum, minimum, average and total overlap between query and page title \\
&$b\_url\_q\_overlap\_\{max,min,avg,total\}$& \{0.16,0.064,0.133,0.044\} & \{0.041,0.018,0.025,0.028\} & Maximum, minimum, average, total term overlap between query and page URL \\
  
\midrule
\multirow{10}{*}{\emph{Mouse}}&$m\_num $  &0.066&0.113& total number of mouseovers in the session \\
&$m\_num\_per\_q $  &0.094&0.053& average number of mouseovers per query \\
&$m\_rank\_max $  &0.091&0.067& max mouseover rank in the session \\
&$m\_rank\_max\_per\_q $  &0.095&0.039& average max mouseover rank per query \\
&$m\_scroll\_dist $  &0.120&0.058& total scroll distance in session \\
&$m\_scroll\_dist\_per\_q $  &0.120&0.025& average scroll distance per query \\
&$m\_scroll\_max\_pos $  &0.142&0.052& max scroll position in session \\
&$m\_scroll\_max\_pos\_per\_q $  &0.127&0.021& average max scroll position per query \\

\bottomrule 
\end{tabular}
\end{adjustbox}
\label{tab:features}
\vspace{-5pt}
\end{table*}

\textbf{Session Features.} The relation between feature $s\_duration$ and different stages of learning has been discussed by Jansen et al.~\cite{jansen2009using}. It has been shown that there is a difference in the duration of sessions among the classifications in Anderson and Krathwohl's taxonomy~\cite{anderson2001taxonomy}. 
White et al.~\cite{white2009characterizing} also found that the sessions conducted by domain experts were generally longer than non-expert sessions. 

\textbf{Query Features. }Several prior works~\cite{jansen2009using, arguello2014predicting, white2009characterizing} have investigated the correlation between query activities in a search session and learning performance. 
Based on the study by White et al.~\cite{white2009characterizing}, the \textit{number of queries} ($q\_num$) applied by experts and non-experts show big differences across domains: non-expert users usually run significantly more queries than experts. 
Jansen et al.~\cite{jansen2009using} also found that the \textit{number of queries} applied on learning tasks classified as \textit{applying stage} was significantly different from other l\textit{earning stages}. 

The \textit{length of queries} ($q\_term\_max\{min,avg,total\}$) has been found to have a strong correlation with learning outcome by Zhang et al.~\cite{zhang2011predicting}. Their study shows that the \textit{average query length} and user domain knowledge is correlated with a Pearson correlation score of 0.344. 

The \textit{complexity of queries} ($q\_complexity\_max\_diff$) has been investigated by Eickhoff et al.~\cite{eickhoff2014lessons}, and has been found to evolve during the learning process. We applied the same query complexity measure as in~\cite{eickhoff2014lessons}, which is computed based on the dictionary created by Kuperman et al.~\cite{kuperman2012age} that contains a listing of more than 30,000 English words along with the age at which native speakers typically learn the term. The maximum age of acquisition across all query terms is used as query complexity. 

Furthermore, the investigation from Arguello~\cite{arguello2014predicting} shows that beside the number of total terms, the \textit{number of unique terms} ($q\_uniq\_term\_\{max, min, avg, total\}$, $q\_uniq\_term\_ratio$) in the session is strongly correlated with knowledge level on the task, while the number and ratio of stop words do not have a big difference when comparing between search sessions with different levels of domain knowledge.

As we aim at predicting knowledge state change during a session, similarly to the features discussed above, we extract the features $q\_len\_\{first, last\}$, $q\_uniq\_term\_\{first,last\}$, which potentially are indicators of the knowledge level at the beginning and end of the session.

\textbf{SERP Activity Features. }Some activities on SERP have also been investigated by previous works. Specifically, Collins-Thompson~\cite{collins2016assessing} found that the \textit{total number of clicks on SERP} ($SERP\_click$) is strongly correlated with a user's understanding of the topic. The analysis shows that users tend to click more often when having stronger interest in the topic. 

The \textit{ranking position} of the clicked URL on SERP has also been shown to be a strong indicator of user domain knowledge by Zhang et al.~\cite{zhang2011predicting}. 
In~\cite{arguello2014predicting}, the authors discovered that the difficult tasks with which a user is less knowledgeable are associated with more clicks ($SERP\_click$), more clicks on lower ranks ($SERP\_click\_rank\_\{highest,lowest,avg\}$), more abandoned queries ($SERP\_no\_click\_query\_\{num,pct\}$), i.e. queries without clicks, longer \textit{time till first click} ($SERP\_avg\_time\_to\_first\_click$) and longer \textit{time till next click} ($SERP\_click\_interval$).

\textbf{Browsing Features. } Browsing features such as \textit{number of documents viewed} ($b\_num$, $b\_uniq\_num$) and \textit{average number of documents viewed per query} ($b\_num\_per\_q$, $b\_uniq\_num\_per\_q$) were shown by several previous works~\cite{eickhoff2014lessons, jansen2009using, arguello2014predicting, gwizdka2006can} to be positively correlated with the knowledge improvement. More detailed features corresponding to the browsing behavior have also been studied, indicating that the more difficult a task is for a user, the higher the \textit{ratio of revisited pages} ($b\_revisited\_ratio$) is. 

Despite the number of pages visited, the time spent (corresponds to features $b\_time\_total$, $b\_time\_\{max,avg\}\_per\_q$ etc.) on the accessed pages are found to vary to a large extent between domain expert and non-expert~\cite{white2009characterizing}. Feature $SERP\_time\_\{total,avg,max\}$ was shown to be effective for predicting the user's assessment of task difficulty~\cite{arguello2014predicting}, which is subject to the user's knowledge state.  

We further distinguish the viewed pages into two sets \{pages navigated through SERP, pages navigated through non-SERP\}, by parsing its ancestor page. Hence we extract the features $b\_\{num,pct\}\_from\_SERP$ and $b\_\{num,pct\}\_from\_non\_SERP$ that are motivated by the features introduced above.

The content of the accessed Web documents strongly influence the user's learning outcome. White et al.~\cite{white2009characterizing} found that domain experts encountered different and more diverse domains (feature $b\_distint\_domain\_num$) than domain novices.
Several other document content related features: \textit{page size ($b\_page\_size\_\{max,min,avg,total\}$), title length $b\_ttl\_len\_\{max,min,avg,total\}$} have also been found to evolve during the learning process~\cite{eickhoff2014lessons}. Based on the assumption that domain experts and novices have different capabilities of choosing learning resources, for instance, experts are able to recognize useful documents without query terms presented in the page title, we computed features based on the overlap between page title and query ($b\_url\_q\_overlap\_\{max,min,avg,total\}$). The page URL ($b\_ttl\_q\_overlap\_\{max,min,avg,total\}$) as a complementary source containing hints about a page's content has also been considered in the feature extraction process.

\textbf{Mouse Features.} Features in the \textit{Mouse} category are indicators of quantity and quality of user interactions with a knowledge source and were also shown to be effective for predicting the user's assessment of task difficulty~\cite{arguello2014predicting}. 

\subsection{Feature Analysis and Selection} \label{sec:feature_selection} 
As a basis for feature selection, we analyze the features with respect to their relationship to knowledge gain and knowledge state, as well as their redundancy.

\textbf{Correlation between feature and KG (KS).} In order to select the most influential features for the prediction task, we compute $Corr(f_i, \Delta k(t_i, t_j))$ and $Corr(f_i, k(t_j) )$, i.e. the Pearson correlation coefficient between each feature and the knowledge gain (knowledge state). The correlation scores are shown in Table \ref{tab:features}. Based on the computed score, we select the features fulfilling the condition $Corr(f_i, \Delta k(t_i, t_j) ) \geq \beta$ for the knowledge gain prediction task and $Corr(f_i, k(t_j) ) \geq \gamma$ for the knowledge state prediction task. Performance of the prediction model using features selected based on varied $\beta$ and $\gamma$ has been evaluated and corresponding results are presented in Section~\ref{sec:evaluation}.

\textbf{Correlation between features.} We compute the Pearson correlation coefficient between each pair of features $Corr(f_i, f_j)$. If $Corr(f_i, f_j) \geq \tau$, i.e. features appear to be not independent, we remove the feature from the feature set, that has lower $Corr(f_i, \Delta k(t_i, t_j) )$ respectively lower $Corr(f_i, k(t_j) )$ for knowledge gain (state) prediction. We evaluate the performance of the prediction model for different values of $\tau$. The feature selection results are reported in Section~\ref{sec:evaluation}.

\section{Evaluation - Experimental Setup} \label{sec:exp}

\subsection{Approach Configurations \& Baselines} \label{sec:config}
\subsubsection{Configurations and Parameters}

\begin{itemize}[leftmargin=*, nosep]

\item \textbf{Classifier.} We apply a range of standard models for the classification of the knowledge gain and knowledge state, namely, Naive Bayes (NB), Logistic Regression (LR), Support Vector Machine (SVM), Random Forrest (RF), and Multilayer Perceptron (MP). For our experiments, we used the Weka library for Java\footnote{https://www.cs.waikato.ac.nz/ml/weka/}. 
For each of the configurations described below, we perform grid search to tune the hyperparameters of all of the classifiers. 
In Section \ref{sec:evaluation}, we report the result of the best performing hyperparameter configuration for each classifier. 

\item \textbf{$\beta$ ($\gamma$)- threshold for feature selection based on correlation between feature and KG (KS).} We compare prediction performance before and after applying the selection based on feature-KG (KS) correlation. We set the threshold $\beta$ ($\gamma$) for selecting the features in the range of \{0.0, 0.05, 0.1, 0.15, 0.2, 0.25, 0.3\} (\{0.0, 0.05, 0.1, 0.15\}). We omit results for larger $\beta$ ($\gamma$), as the resulting number of features is insufficient for training a classifier.

\item \textbf{$\tau$ - threshold for feature selection based on correlation between features.} We also experimented with different $\tau$ in the range of \{1.0, 0.95, 0.9, 0.85, 0.8, 0.75, 0.7\}. The number of features in the feature set corresponding to given $\tau$, $\beta$ and $\gamma$ is reported in Table \ref{tab:feature_num}.
\end{itemize}

\begin{table}[h]
\centering
\vspace{-5pt}
\caption{\small{Number of features of different configurations.}}
\vspace{-10pt}
\begin{adjustbox}{max width=0.45\textwidth}
\small
\begin{tabular}{ p{0.9cm} | p{0.4cm}  p{0.4cm}p{0.4cm}p{0.4cm}p{0.4cm}p{0.4cm}p{0.4cm}| p{0.4cm}p{0.4cm}p{0.4cm}p{0.4cm} }
\toprule
&\multicolumn{7}{c|}{$\beta$ (\textbf{KG}) } & \multicolumn{4}{c}{$\gamma$ (\textbf{KS}) } \\
& 0.0 & 0.05 & 0.1 &0.15 & 0.2 &0.25 &0.3 & 0.0 & 0.05 & 0.1 & 0.15 \\ \midrule
$\tau$ = 1.0 & 70 & 43 & 16 & 6 & 4 & 2 & 1 & 70 & 41 & 11 & 0  \\
$\tau$ = 0.95 &66 & 42 & 16 & 6 & 4 & 2 & 1 & 66 & 38 & 11 & 0\\
$\tau$ = 0.9 & 56 & 37 & 13 & 6 & 4 & 2 & 1 & 58 & 34 & 11 & 0\\
$\tau$ = 0.85 & 43 & 29 & 10 & 6 & 4 & 2 & 1 & 44 & 24 & 9 & 0 \\
$\tau$ = 0.8 & 39 & 26 & 7 & 4 & 2 & 1 & 1& 38 & 19 & 8 & 0 \\
$\tau$ = 0.75 & 37 & 25 & 7 & 4 & 2 & 1 & 1 & 34 & 17 & 7 & 0\\
$\tau$ = 0.7 & 33 & 24 & 6 & 3 & 1 & 1 & 1 & 32 & 16 & 7 & 0\\
\bottomrule 
\end{tabular}
\end{adjustbox}
\label{tab:feature_num}
\vspace{-15pt}
\end{table}

\subsubsection{Baseline}
As discussed in Section \ref{sec:sota}, the tasks addressed in this paper are comparably novel. To the best of our knowledge, there are no existing baselines for the task of knowledge gain prediction during informational Web search missions. Therefore, we compare our approach for a number of configurations (described above), using multiple standard classification models. For the prediction of knowledge state $k(t_j)$, we compare our approach in addition to one existing baseline ~\cite{zhang2015predicting}. 
\textbf{$KS_{Zhang}$} refers to the linear regression model fitted by Zhang et al.~\cite{zhang2015predicting} for domain knowledge prediction as shown in Equation \ref{eq:zhang_predicting}.
\vspace{-10pt}

\begin{small}
\begin{equation} \label{eq:zhang_predicting}
\vspace{-10pt}
KS_{Zhang}=-1.466 + 0.039\cdot Saved + 0.147 \cdot Q_{len} + 0.130 \cdot Rel_{mean}
\end{equation}
\end{small}

\textit{Saved} represents the number of documents saved by the user, which is an extremely sparse feature in a real search environment and does not appear in our dataset. \textit{$Q_{len}$} is the mean query length and \textit{$Rel_{mean}$} is the mean rank of documents opened in SERPs. As the output of the baseline regression model is a real number, we convert the result into 3 classes according to the definition given in Section~\ref{sec:data_grouping}.

\subsection{Evaluation Metrics}

For both tasks, we run repeated 10-fold cross-validation with 10 repetitions on all the approaches and configurations described in Section \ref{sec:config} and evaluate the results according to the following metrics:

\begin{itemize}[leftmargin=*, nosep]
\item \textbf{Accuracy ($Accu$) across all classes:} percentage of search sessions that were classified with the correct class label.
\item \textbf{Precision ($P$), Recall ($R$), F1 ($F1$) score of class $i$:} we compute the standard precision, recall and F1 score on the prediction result of each class $i$.
\item \textbf{Macro average of precision ($P$), recall ($R$), and F1 ($F1$):} the average of the corresponding score across 3 classes.
\item \textbf{Runtime:} the time consumed for completing the 10-fold cross-validation on experimental dataset in milliseconds.
\end{itemize}

To analyze the usefulness of individual features, we make use of the \textit{Mean Decrease Accuracy (MDA)} metric, which is based on the Random Forest model, i.e. a very well performing model for both tasks as shown in Section~\ref{sec:evaluation}. MDA quantifies the importance of a feature by measuring the change in prediction accuracy of the Random Forest model when the values of the feature are randomly permuted compared to the original observations.

\begin{table*}[!htbp]
\vspace{-5pt}
	\caption{Performance in knowledge gain prediction task.}
\vspace{-10pt}
	\label{tab:kg_result}
	\small
	\centering
 	\scalebox{1}{
	\begin{tabular}{p{1cm}| p{0.5cm}p{0.5cm} | p{1cm}p{1cm}|p{0.5cm}p{0.5cm}p{0.5cm}|p{0.5cm}p{0.5cm}p{0.5cm}|p{0.5cm}p{0.5cm}p{0.5cm}|p{0.5cm}p{0.5cm}p{0.5cm}|p{0.5cm}p{0.5cm}p{0.5cm}|p{0.5cm}}
	\toprule
    & & & & & \multicolumn{3}{c}{\textbf{Low}}& \multicolumn{3}{c}{\textbf{Moderate}}& \multicolumn{3}{c}{\textbf{High}}& \multicolumn{3}{c}{\textbf{Macro average}} & \textbf{All} \\
   \textbf{Method} & $\tau$ &  $\beta$ & \textbf{\#Features} & \textbf{Runtime} & \textbf{P}& \textbf{R}& \textbf{F1}& \textbf{P}& \textbf{R}& \textbf{F1}& \textbf{P}& \textbf{R}& \textbf{F1}& \textbf{P}& \textbf{R}& \textbf{F1} &\textbf{Accu} \\
   \toprule
   
\multirow{ 1}{*}{NB} & $\geq$0.85 & 0.25 & 2 &  19.1 &  0.450 &  0.747 &  0.562 &  0.483 &  0.268 &  0.344 &  0.513 &  0.384 &  0.439 &  0.482 &  0.467 &  0.448 &  0.469 \\

\multirow{ 1}{*}{LR}& 0.85 & 0.05 & 29 &  653.9 &  0.498 &  0.537 &  0.516 &  0.459 &  0.382 &  0.416 &  0.379 &  0.431 &  0.403 &  0.445 &  0.450 &  0.445 &  0.450 \\

\multirow{ 1}{*}{SVM}&  0.90  &  0.00  &  56  &  441.6 &  0.488 &  0.595 &  0.536 &  0.487 &  0.340 &  0.400 &  0.410 &  0.469 &  0.437 &  0.462 &  0.468 &  0.458 &  0.465\\

\multirow{ 1}{*}{RF} &  0.95  &  0.00  &  66  &  3739.3 &  0.521 &  0.542 &  0.531 &  0.469 &  0.410 &  0.437 &  0.425 &  0.480 &  0.450 &  0.472 &  0.477 &  \textbf{0.473} &  \textbf{0.475}\\

 \multirow{ 1}{*}{MP}  & $\geq$0.85 & 0.25 & 2  &  1919.3 &  0.452 &  0.556 &  0.497 &  0.421 &  0.312 &  0.356 &  0.425 &  0.450 &  0.435 &  0.433 &  0.439 &  0.429 &  0.435\\
   \bottomrule
	\end{tabular}}
 %   \vspace{-5pt}
\end{table*}

\begin{figure*}[!htbp]
\centering
\includegraphics[clip=true, trim=2pt 0pt 4pt 4pt, width=0.98\textwidth]{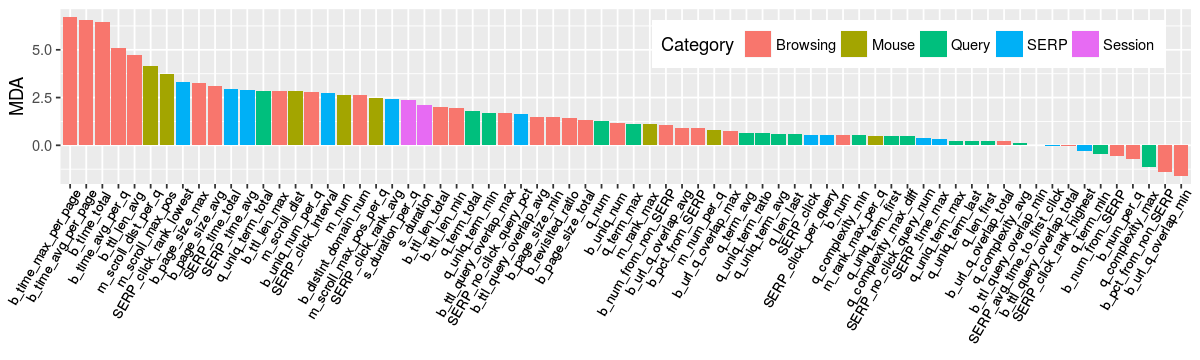}
\caption{Feature importance for knowledge gain prediction.}
\label{fig:f_imp_kg}
\end{figure*}

\section{Results: Prediction Performance and Feature Analysis} \label{sec:evaluation}
In this section, we report the evaluation results of the prediction performance as well as an analysis of feature importance.

\subsection{Knowledge Gain Prediction} \label{sec:eva_kg}

\textbf{Performance of different Configurations.} For each of the 245 distinct configurations described in Section \ref{sec:exp}, we run repeated cross-validation as described in the previous section.

From all the different combinations of $\tau$ and $\beta$ as listed in Table \ref{tab:feature_num}, we present the result of the configuration that produces the highest accuracy for each classifier in Table \ref{tab:kg_result} due to space constraints. A complete set of the evaluation results are available online\footnote{https://sites.google.com/view/predicting-user-knowledge}. We observed that in the knowledge gain prediction task, the highest average F1 score across classes and the highest accuracy always appear in the same configuration for all the classifiers except for LR, where there is a minor difference of .001 in F1 between the two.

The \textit{Random Forest} classifier achieves the best performance in terms of accuracy and average F1 score, slightly outperforming Naive Bayes and SVM. As shown, Naive Bayes is the most efficient classifier in terms of computation time for feature sets of comparable size. 
Comparing classification performance for the different classes, the results for each of the classifiers consistently show higher F1 scores for the `Low' and `High' knowledge gain classes than for the `Moderate' knowledge gain class. Overall, prediction accuracy is above 0.435 and the F1 score is above 0.429 for all of the classifiers, which indicates that the set of features we extracted from search activities can provide meaningful evidence for predicting knowledge gain. 

\textbf{Feature Impact.} The MDA results of each feature are shown in Figure \ref{fig:f_imp_kg}. 
Based on the result, the most important features are: $b\_time\_max\_per\_page$, $b\_time\_avg\_per\_page$ and $b\_time\_total$. Mostly active time related features which reflect the effort users spend on learning. 
Similarly, these features are immediately followed by $m\_scroll\_dist\_per\_q$, $m\_scroll\_max\_pos$, and $SERP\_click\_rank\_lowest$, three features that are indicators for the amount of information a user has seen during the session.

As shown in Table \ref{tab:kg_result}, the NB and MP classifier performs best when using 2 features only, namely $b\_time\_max\_per\_page$ and $b\_time\_avg\_per\_page$. These two features are confirmed as the most important features by our feature importance analysis.% (Figure \ref{fig:f_imp_kg}). 

Regarding feature categories, the 10 most useful features in terms of MDA belong to the browsing, mouse and SERP categories. 
Surprisingly, although multiple query features had above average correlation to knowledge gain (see Table~\ref{tab:features}) -- in particular, the features related to query complexity ($q\_complexity\_\{max,min,avg\}$ and $q\_complexity\_max\_diff$) had correlations ranging from .086 to .097, compared to the median of .042 -- $q\_uniq\_term\_total$ is the only query feature among the 25 highest ranked according to MDA. Analogously, both session features $s\_duration$ and $s\_duration\_per\_q$ appear among the 25 highest ranked features despite their relatively low correlations of -.02 and -.019.

\subsection{Knowledge State Prediction}

\begin{table*}[!htbp]
	\caption{Performance in knowledge state prediction task.}
    \vspace{-10pt}
	\label{tab:ks_result}
	\small
	\centering
 	\scalebox{1}{
	\begin{tabular}{p{1cm}| p{0.5cm}p{0.5cm} | p{1cm}p{1cm}|p{0.5cm}p{0.5cm}p{0.5cm}|p{0.5cm}p{0.5cm}p{0.5cm}|p{0.5cm}p{0.5cm}p{0.5cm}|p{0.5cm}p{0.5cm}p{0.5cm}|p{0.5cm}p{0.5cm}p{0.5cm}|p{0.5cm}}
	\toprule
    & & & & & \multicolumn{3}{c}{\textbf{Low}}& \multicolumn{3}{c}{\textbf{Moderate}}& \multicolumn{3}{c}{\textbf{High}}& \multicolumn{3}{c}{\textbf{Macro average}} & \textbf{All} \\
   \textbf{Method} & $\tau$ &  $\gamma$ & \textbf{\#Features} & \textbf{Runtime} & \textbf{P}& \textbf{R}& \textbf{F1}& \textbf{P}& \textbf{R}& \textbf{F1}& \textbf{P}& \textbf{R}& \textbf{F1}& \textbf{P}& \textbf{R}& \textbf{F1} &\textbf{Accu} \\
   \toprule
\multirow{ 1}{*}{NB} & $\leq$0.75 & 0.1 & 7 &  23.5 &  0.352 &  0.712 &  0.470 &  0.424 &  0.218 &  0.287 &  0.370 &  0.211 &  0.268 &  0.382 &  0.380 &  0.342 &  0.369\\

\multirow{ 1}{*}{LR} &  1.00  &  0.05  &  41  &  797.7 &  0.338 &  0.383 &  0.359 &  0.402 &  0.368 &  0.384 &  0.372 &  0.359 &  0.366 &  0.370 &  0.370 &  0.370 &  0.370\\

\multirow{ 1}{*}{SVM}& 0.95  &  0.05  &  38  &  292.3 &  0.359 &  0.479 &  0.409 &  0.395 &  0.303 &  0.342 &  0.409 &  0.386 &  0.397 &  0.388 &  0.389 &  0.383 &  0.385\\

\multirow{ 1}{*}{RF} &  1.00  &  0.00  &  70  &  4023.4 &  0.443 &  0.456 &  0.449 &  0.394 &  0.358 &  0.374 &  0.418 &  0.447 &  0.432 &  0.418 &  0.421 &  \textbf{0.418} &  \textbf{0.418}\\

 \multirow{ 1}{*}{MP} & 1.00 & 0.05 & 41 & 43619 & 0.380 & 0.414 & 0.396 & 0.398 & 0.298 & 0.341 & 0.385 & 0.461 & 0.419 & 0.388 & 0.391 & 0.385 & 0.387\\
 
 \midrule
 $KS_{Zhang}$ & - & - & 2 & 23 & 0.320 & 0.428 & 0.366 & 0.328 & 0.240 & 0.277 & 0.362 & 0.355 & 0.359 & 0.337 & 0.341 & 0.334 & 0.335\\
   \bottomrule
	\end{tabular}}
\end{table*}

\begin{figure*}[h]
\centering
\includegraphics[clip=true, trim=2pt 0pt 4pt 4pt, width=0.98\textwidth]{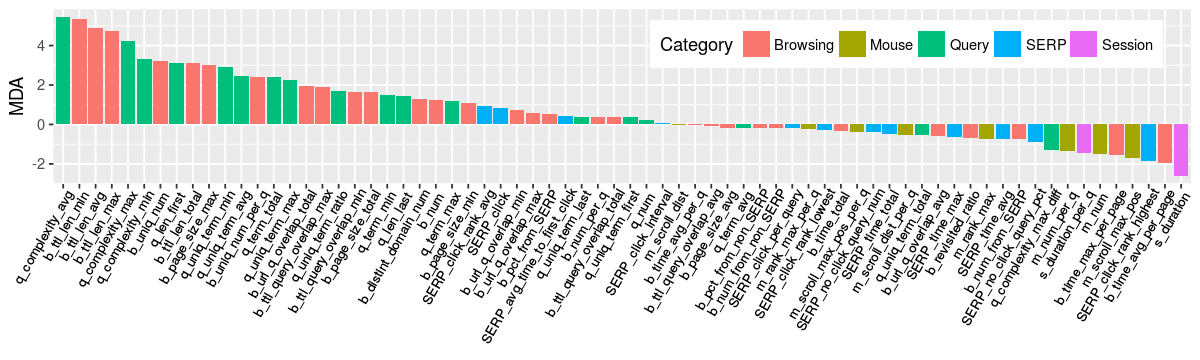}
\caption{Feature importance for knowledge state prediction.}
\label{fig:f_imp_ks}
\end{figure*}

\textbf{Performance of different Configurations.} We have experimented with all different combinations of $\tau$ and $\gamma$ as listed in Table \ref{tab:feature_num} for all considered classifiers. The result of the configuration that produces the highest accuracy for each classifier is shown in Table \ref{tab:ks_result}. We observe that in the knowledge state prediction task, the highest average F1 score across classes and the highest accuracy always appear in the same configuration for all the classifiers except Naive Bayes (average F1 of the highest accuracy configuration is 0.006 lower than the maximum average F1). 

Among all evaluated classifiers, Random Forest reaches the highest accuracy and F1 score, outperforming the other classifiers.

\textbf{Comparison to Baseline.} We compare the performance of our approach against the baseline method ($KS_{Zhang}$), shown in the last row in Table \ref{tab:ks_result}. The result suggests that, the linear regression model fitted in previous work based on data collected through a lab study does not perform well in the knowledge state prediction task and is outperformed by all five classifiers following our approach.

\textbf{Feature Impact.} The MDA results of each feature in the knowledge state prediction tasks are shown in Figure \ref{fig:f_imp_ks}. 
The most important features ($q\_complexity\_avg$, $b\_ttl\_len\_min$ and $b\_ttl\_len\_avg$) reflect the user's capability of constructing a query and choosing relevant resources. In terms of feature categories, all of the highest ranked features for this task belong to the query and browsing categories. 

Compared to the knowledge gain task, query complexity features ($q\_complexity\_\{min,max,avg\}$) are considerably more useful, while features related to time and effort invested, like $b\_time\_max\_per\_page$ and $b\_time\_avg\_per\_page$, are among the lowest ranked. 
Other query features related to the used vocabulary (e.g. $q\_uniq\_term\_min$, $q\_uniq\_term\_avg$, and $q\_term\_total$) are ranked similarly highly. 
Apparently, while the time taken by users to take in the discovered documents is predictive of their knowledge gain, their capability of using complex queries and selecting relevant resources reveals more about their knowledge state.

\section{Discussion}

Based on the experimental results, we conclude that: i) knowledge gain (state) can be predicted during informational search sessions with a certain level of accuracy, ii) performance of the knowledge gain prediction appears to be generally better, suggesting that the task is easier given the nature of our data, and iii) the performance of the prediction approach is better for more extreme classes, i.e. for low and high knowledge gain (state) classes, whereas performance on the moderate classes is lowest in both tasks, presumably due to the moderate classes being the most overlapping ones with respect to their characteristics. In this section we discuss some of the reasons behind these observations.

Most of the features we considered were found to correlate rather weakly with knowledge gain (state). Intuitively, this could be due to the limited duration of the search sessions (just over 5 minutes on average). This could potentially reduce the predictive power of certain features, such as the number of queries or the number of accessed documents. This also rendered evolution-oriented features, which would capture the evolution of queries and behavior throughout a session predictively poor. While these would supposedly be highly indicative of the knowledge gain, they require longer sessions than are usually observable in real-world search sessions as well as in our experimental data.

For the prediction of knowledge gain, our feature analysis result shows that the most important features are the ones related to the user's active time. As our experimental dataset contains mostly short sessions, it is understandable that the time spent affects the knowledge gain strongly. However, we believe that in longer search sessions, the learning pattern and the initial knowledge state of a user might be more influential for the knowledge gain than in short sessions. Further experiments are required to establish this.

The results suggest that with the presented approach, the knowledge gain prediction is an easier task than the knowledge state prediction. As shown in Figure \ref{fig:f_imp_ks}, the most important features for knowledge state prediction are the features related to the content of queries and browsed documents. 
Intuitively, these features are also central to the knowledge gain prediction task. Yet, we observe that although the topic descriptions that were given to the users typically provided central keywords for the first query, only a very limited set of queries (1-2) are fired by most users. Given the small number of queries in each session, the query features are less distinguishable and hence, less indicative of the knowledge gain. Thus, query evolution is observable only to a very limited extent.

\section{Conclusions and Future Work}
In this paper, we propose to use classification models to predict user knowledge state and knowledge gain from behavioral data captured during real-world informational search sessions. Given the lack of available datasets and ground truths, we have created an experimental dataset using crowdsourcing, capturing 468 informational search sessions including user interactions and behavioral traces throughout the process, together with calibration and corresponding post-session knowledge test results. 

Previous work related to the aforementioned prediction tasks is still very limited. However, existing state of the art was considered when identifying a novel set of 70 features, partially motivated from related work as well as from exploring our gathered data. Classification experiments were conducted with 5 classification models in a variety of configurations and in comparison to a baseline in the knowledge state prediction task.

The experimental results underline that a user's knowledge gain and knowledge state can be modeled based on a user's online interactions observable throughout the search process. Through feature analysis, we provide evidence for an improved understanding between individual user behavior and the corresponding knowledge state and change. Alongside these results, we also make the gathered dataset available. This dataset captures user interactions throughout diverse informational search sessions and corresponding knowledge assessments, and thereby provides a resource which can facilitate further research in this area.

As a part of future work, we aim to reproduce and refine the findings in more varied search sessions, where durations and learning intents are more diverse; involving considerably longer search sessions and, for instance, procedural knowledge rather than intents focused on declarative knowledge only. This would provide the opportunity to observe evolution-oriented features, such as considering the evolution of queries, their length and complexity. 

Potential applications of this work include the consideration of user knowledge and the expected learning progress of a user as part of Web search engines and information retrieval approaches, or within informal learning-oriented search settings, such as libraries or knowledge- and resource-centric online platforms.

\balance
\bibliographystyle{abbrv}
%\newpage
\bibliography{references}  

\end{document}